\documentclass{article}[12pt] \usepackage{latexsym}
\usepackage{amsfonts} \usepackage{amsmath,cite,epsfig,graphicx} 

\newcommand{\kt}[1]{\ensuremath{|#1\rangle}}
\newcommand{\br}[1]{\ensuremath {\langle #1|}}
\newcommand{\bk}[2]{\ensuremath {\langle #1|#2 \rangle}}
\newcommand{\wig}{\mathbf{p},\xi [\sm,j]}

\newcommand{\sm}{\mathsf{s}}
\newcommand{\ms}{\mathsf{s}}
\newcommand{\HS}{{\mathcal{H}}}
\newcommand{\Phx}{\Phi^\times}
\newcommand{\vp}{{\bf p}}

\begin{document}

\title{Poincar\'e Semigroup Symmetry as an Emergent Property of Unstable Systems}

\author{N.L.~Harshman\\Department of Computer Science, Audio Technology, and Physics\\
American University\\ Washington, DC 20016}

\maketitle

\vspace{1cm}

Abstract: 
\normalsize

The notion that elementary systems correspond to irreducible representations of the Poincar\'e group is the starting point for this paper, which then goes on to discuss how a semigroup for the time evolution of unstable states and resonances could emerge from the underlying Poincar\'e symmetry.  Important tools in this analysis are the Clebsch-Gordan coefficients for the Poincar\'e group.

\section{Introduction}

One productive perspective on theoretical particle physics is Wigner's notion of the stable particle as a unitary, irreducible representation (UIR) of the Poincar\'e group.  However, this idea loses some of its clarity when applied to stable particles that are not elementary, such as the proton, made of its three quarks and a sea of gluons and quark-antiquark pairs, or even the electron, with its cloud of virtual photons.  Trying to fit unstable particles into the UIR scheme creates even more interesting questions, and is the subject of this article.

Before proceeding, I must explain my terminology, which I will introduce with this edited quotation from a famous article by Newton and Wigner\cite{newtonwigner}.  This quotation also touches on some of the main ideas of this paper and therefore serves as a good introduction.

\begin{quotation}
It is well known that invariance arguments suffice to obtain an enumeration of the relativistic equations for elementary systems.  The concept of an ``elementary system'' is, however, not quite identical with the intuitive concept of an elementary particle...The definition under which the aforementioned enumeration can be made is...: it requires that all states of the system be obtainable from the relativistic transforms of any state by superpositions....Every system, even one consisting of an arbitrary number of particles can be decomposed into elementary systems.
\end{quotation}

The first two sentences make the distinction between elementary systems and elementary particles.  In what follows, stable particles will be considered as elementary systems, even if they are not elementary particles.  The third sentence is a serviceable definition of an irreducible representation, therefore, I will use the terms UIR, elementary system, and stable particle interchangeably as context suggests.  It is the fourth and final sentence that makes a truly bold claim that will be explored but not resolved in this article.  Mathematically, the statement suggests that every particle configuration should be expressible as direct products of UIRs of the Poincar\'e group.  In other words, the state of any particle or particles (and even unstable particles) should be expressible in a basis constructed of stable particles.  Just what this means, how such constructions are made, and how it connects to questions of time symmetry and asymptotic completeness will be considered in this paper after I give  brief perspective on the unstable particle zoo.

Before continuing though, I must justify the title.  As a working definition of emergent properties, I mean such properties arise out of more fundamental entities and yet are novel or irreducible with respect to them.  In physical systems, emergent properties tend to develop as the number of constituent systems or complexity of those systems grows.  There are many examples of the productivity of this concept in science from biology to adaptive computer networks.  The point of view explored in this paper is that the characteristic behavior of unstable particles is expressed by the Poincar\'e semigroup, which allows only space-time translations into the forward light cone, and this is just such an emergent property.  Unstable states at a variety of time and energy scales (not just in particle physics) evidence two features: exponential decay and Breit-Wigner (or Lorentzian) resonance amplitudes.  By looking at how instability is included in quantum theory and how stable particles are represented, I hope to raise interesting questions about how this ubiquitous property emerges.

\section{Characterizing Particles}

A truly satisfying particle theory would start with a very small set (or even an empty set) of physical parameters and from them be able to predict and/or explain the characteristic parameters of all particles.  By characteristic parameters, as an example I mean the data listed in the ``Review of Particle Physics''~\cite{PDG} which is used to distinguish between different particles. Although the Standard Model does a good job at many things, it is not yet this satisfactory theory and instead the characteristic parameters must be treated in various ways.  For some of the characteristic parameters there is no theory that predicts their values; they are input parameters into the Standard Model like the lepton masses.  Some other characteristic parameters are related by calculations within the Standard Model to the fundamental parameters.  Others are related in principle by the theory, but the the calculations cannot be evaluated to a level of accuracy sufficient for comparison with experimental data, such as the hadron masses, and must be treated as independent phenomenological parameters or by phenomenological theories.  In any case, we can group the parameters into three rough groups based on their physical nature.

The first set of parameters are intimately related to the symmetries of space-time.  These are the mass $m$ and spin $j$ of the particles and the intrinsic parity.  For unstable particles, it is generally accepted that the finite lifetime leads to an uncertainty in mass as evidenced by the shape of resonant cross sections; recently Blum and Saller have raised the question whether a similar uncertainty also exists for spin~\cite{blumsaller}.  These space-time characteristic parameters fit nicely into the Wigner UIR particle picture, to be discussed in more detail later.

Another set of internal parameters such as charge, lepton number and flavor may also be seen as consequences of some internal symmetry group or may just have to be arrived at phenomenologically.  For non-elementary particles, some explanations rely on the properties of constituent particles.  For composite particles, other characteristic information may take the form of form factors and other structure functions.  

Finally, if the particle is unstable, it will have characteristic instability parameters, most notably lifetime or width, but also branching ratios and perhaps other information like phase relations between decay channels.  I will focus the rest of this section on lifetime and width, since it is these parameters that are associated with the Poincar\'e semigroup and time asymmetry.

Let us look at the spectrum of particles.  Figure 1, taken from \cite{ajp03}, depicts the mass $M$ and width $\Gamma$ of 139 unstable
particles, with mass plotted logarithmically on the horizontal
axis and width plotted logarithmically vertically.  The shape of
each plotted point indicates the type of particle it is (gauge boson,
e.g.)
and for the hadrons the style  
indicates some information about the quark content.  

\begin{figure}
\includegraphics[scale=0.7]{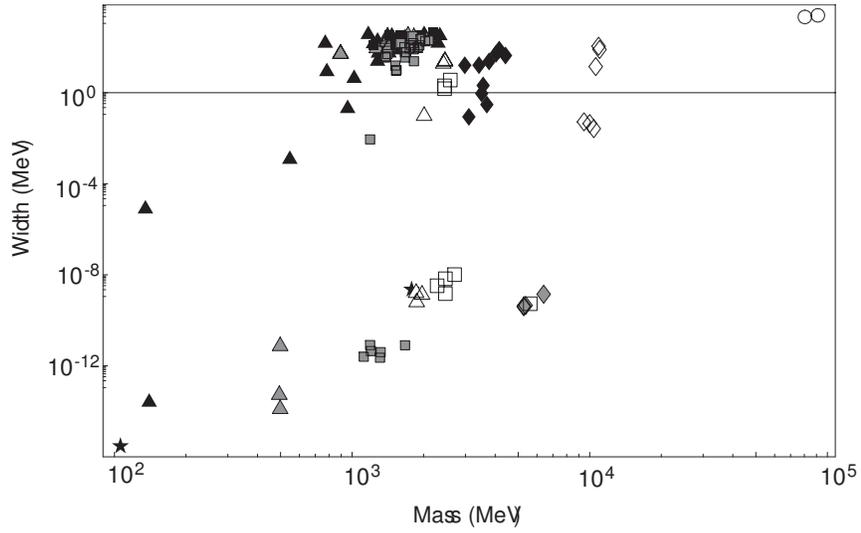}
\caption{Log-log plot of Mass/MeV versus Width/MeV.  Choice
of 139 unstable
particles plotted described in text.  Key:
hollow circles---gauge bosons; black stars---leptons; black triangles---light unflavored mesons; gray triangles---strange
mesons; hollow triangles---flavored charmed mesons (including
charmed/strange mesons); black diamonds---unflavored charmed
mesons; gray diamonds---flavored bottom mesons (including 
bottom/strange and bottom/charmed
mesons); hollow diamonds---unflavored bottom mesons; black squares---$N$ and $\Delta$ baryons; gray squares---strange baryons (including $\Lambda$, $\Sigma$, $\Xi$ and
$\Omega$ baryons); hollow squares---charmed and bottom baryons.}
\end{figure}

The data for the 139 unstable particles come from the 2002 edition of
\emph{The Review of 
Particle Physics}~\cite{PDG}, and in particular the list of well-known,
reasonably well-measured
unstable particles from the ``Summary Tables of Particle Properties''
therein.  Not every particle in the ``Summary Table'' has been
included; only those particles found in the file \cite{PDGMC} that the Particle
Data Group tabulates, of the mass and width data,
for use in
Monte Carlo event generators and detector simulators.  For unstable
particles whose lifetimes $\tau$ are quoted in the
\emph{Review}~\cite{PDG} and not their widths,
the width values in the Monte Carlo file are found using
the Weisskopf-Wigner relation $\Gamma = \hbar/\tau$ (more will be said
about this later).

The list of particles from the Monte Carlo file \cite{PDGMC} has been
modified and applied in the following way in Fig.~1:
\begin{enumerate}
\item The stable particles, the proton, electron,
photon and neutrinos, are excluded.
\item The nearly-stable neutron is neglected for reasons of scale.
\item The Monte Carlo
file includes some particles for which only an upper bound of the
width has been measured.  They 
have been excluded.  Examples include some light
unflavored meson resonances like the $f_0(980)$, other meson
resonances such as the $D_s^{*\pm}$ and $\chi_{b0}(1P)$, and a few
baryon resonances like the $\Sigma_c(2520)^+$.
\item The top quark, not truly an independent particle like the others
in the list, is not included.
\item The symbol plotted for a particle also represents its
antiparticle, except for the neutral K-mesons.  For these, the mass
eigenstates $K^0_S$ and $K^0_L$ are plotted instead of
the flavor eigenstates $K^0$ and $\bar{K}^0$.

\item A single symbol is plotted for all different charge-species of a
hadron unless different masses for different charges have
been measured.  For example, each point representing a $\Delta$ baryon
represents all four charge species $\{++, +, 0, -\}$ corresponding to
quark contents $\{uuu, uud, udd, ddd\}$.

\end{enumerate}

Then what unstable particles \emph{are} included?  The weak gauge bosons W
and Z are at the high energy extreme and 
the muon is at the low energy extreme.  The other unstable lepton, the
tau, is in 
the middle, 
along with a host of hadrons made up of five out of the six quarks:
up, down, 
strange, charm and bottom.  While the gauge bosons and leptons are
to our best knowledge structureless; the hadrons are composite.
Subsequent references to particles refer just to this set of
well-established, well-measured unstable particles, and therefore should
not be taken to refer to all possible particles that have or have not
been observed or theorized.

How are the instability parameters of these particles measured?  The answer is very different depending on whether they are observed as decaying states or as resonances.  For decaying states, the lifetime is measured by the exponential decay rate, observed, for example, in an experiment where an ensemble of systems are produced at a well-localized locations.  Then if the decay vertex can be identified, the distance between production and decay can be converted into time with additional kinematic information and the resulting histogram can be fit to an exponential.  For the lifetime to be measurable requires that the lifetime not be too short; the shortest that has been directly measured is the $\pi^0$ with a lifetime of $(8.4 \pm 0.6) \times 10^{-17}$ s, where an exponential was fit to three points~\cite{atherton}.  This corresponds via the Weisskopf-Wigner relation to a width of around $10^{-5}$ MeV.  All decaying states for which the lifetime is measurable decay via the weak interaction and are along the bottom of Fig.~1.

On the other hand, particle resonances are detected as rapid variations (usually peaks) in the
cross section which have a maximum value and a width.  As the 
center-of-mass energy of a collision is scanned over some range, there
may appear an enhancement of the elastic cross section 
or the cross section into a particular set of inelastic channels.
After extracting the background and accounting for uncertainties in
the preparation and detection apparatuses and other effects (such as
radiative corrections), the resonant cross section $\sigma_R$ as a function of
center-of-mass energy (or center-of-mass energy squared $\sm$) can be
extracted.  This process can become more complicated if there are
multiple resonances in the same energy region, interfering resonances,
or background-resonance interference.  With current experimental energy resolutions, a width must be larger than $1-10$ MeV to be directly observed, corresponding to lifetimes shorter than about $10^{-23}$ s.  As a result, there are no particles for which both the lifetime and width can be independently measured and therefore the Weisskopf-Wigner relation has not actually ever been tested.  There are also some particles for which neither the lifetime or width has been directly measured and other techniques have been applied.

We return now to the question raised in the introduction.  Can these particles, which may be elementary, as in the case of the muon, or composite, like hadrons, be decomposed into elementary systems that are irreducible with respect to the Poincar\'e group symmetry?  And if so, how?  Finally, how does the semigroup behavior of their time evolution emerge?  If we believe that the behavior of the resonances and decaying states are not qualitatively different, but only quantitatively different in their different time and energy scales, how is the Weisskopf-Wigner relation (untestable as it currently is) built into the theory?

\section{Theory of Decaying States}

First we consider what a satisfactory theory of decaying states would entail to describe the phenomenology of lifetime measurements.  We want something like
\begin{equation}
|\bk{\rm{decay}\ \rm{products}}{\rm{decaying}\ \rm{state}(t)}|^2 = e^{-t/\tau} = e^{-\Gamma t},\label{decay}
\end{equation}
with the restriction that the equation only holds for $t\geq t_0$, where $t_0$ is the creation time, which is usually well-defined on the timescale given by the lifetime $\tau$.

Just such decaying state vectors were introduced by Gamow a long time ago for analysis of alpha decay~\cite{gamow}.  However, despite the obvious phenomenological utility of such states, there are mathematical problems with their use in standard quantum theory.
\begin{itemize}
\item Hermitian operators like the Hamiltonian have real eigenvalues in the Hilbert space, whereas (\ref{decay}) requires the decaying state to have a complex energy $E - i\Gamma/2$ or some relativistic generalization.
\item Khalfin~\cite{khalfin} proved that there can be no exponentially decaying time evolution for Hilbert space states, although there are Hilbert vectors with time evolution infinitesimally close to exponential decay.
\item Hegerfeldt~\cite{hegerfeldt} proved that time localization like that implied by the time $t_0$ restriction is not possible in the Hilbert space; there will either always be a probability for decay (including for times before creation) or there can be no decay~\cite{pra02}.
\end{itemize}

These are all serious objections if we hold that the quantum theory must be constrained to the Hilbert space.  Physicists have seen fit to leave the Hilbert space before.  The Dirac ket, the plane wave scattering states with delta-function normalization, are not in the Hilbert space but they nonetheless provide such a useful tool that generations of physicists have adopted them in their calculations.

Dirac kets, and Gamow kets, can be put on a firm mathematical footing by invoking the program of Gel'fand triplets or rigged Hilbert spaces (RHS)~\cite{bohmrhs}.  The state vectors (or choosing a specific representation, the wave functions) are restricted to a linear, topological space $\Phi$ with a stronger topology than that of the Hilbert space $\mathcal{H}$ with Lebesgue norm.  For a stronger topology, a suitable and sufficient choice for many systems is that the algebra of observables be continuous on the space.  An example of a such a space of well-behaved vectors for the harmonic oscillator is the Schwartz space of infinitely differentiable, rapidly decreasing functions. The dual space to $\Phi$, $\Phi^\times$, which has a weaker topology than 
$\mathcal{H}$, will, for suitable choice of $\Phi$, have the eigenkets of the algebra of observables, even if they have an unbounded spectrum in $\mathcal{H}$.  For example, the dual of the Schwartz space construction for the harmonic oscillator, the space of tempered distributions, contains the eigenkets of position and momentum.  This triplet of spaces
\begin{equation}
\Phi\subset\mathcal{H}\subset\Phi^\times
\end{equation}
is called a Gel'fand triplet or rigged Hilbert space and is constructed for representations of the algebra of observables relevant for a particular system.

To make the Gamow kets mathematically viable for unstable relativistic particles, a suitable choice for the space $\Phi$ must make the vectors \emph{very} well-behaved so that analytic continuation of energy (or center-of-mass energy squared) is be possible~\cite{rgv}.  A little bit more detail will be provided below, but the kernal of this approach is the Hardy class hypothesis. Instead of using the Hilbert space 
axiom
\begin{equation}\label{sqm}
\{\mbox{space of prepared states}\} = \{\mbox{space of detected observables}\}
=\HS
\end{equation}
or the slightly more general revision
\begin{equation}\label{ssqm}
\{\mbox{space of prepared states}\} = \{\mbox{space of detected observables}\}
=\Phi\subset\HS,
\end{equation}
we distinguish mathematically
between states and observables and make the new
hypothesis~\cite{rhsrev}:
\begin{eqnarray}\label{ourbc}
\mbox{The prepared states are described by:}\
&\{\phi^+\}&=\Phi_-\subset\HS\subset\Phx_-\\
\mbox{and the registered observables by:}\
&\{\psi^-\}&=\Phi_+\subset\HS\subset\Phx_+.\nonumber
\end{eqnarray}
Here we use a pair of
Rigged Hilbert spaces of Hardy type,
where the energy wave functions of the vectors $\phi^+\in\Phi_-$ and 
$\psi^-\in\Phi_+$ are well-behaved Hardy functions in the lower and upper half
complex planes, respectively.  
The Gamow vectors, together with the out-plane wave solutions of the
Lippmann-Schwinger equations, are elements of the space $\Phx_+$.

Choosing different spaces for in-states and out-observables calls into question the notion of asymptotic completeness often invoked in quantum field theory.  If we still hold that any particle configuration can be decomposed into elementary systems at any time, we now must consider whether the elementary systems satisfy the boundary conditions implied by $\Phi_+$ or $\Phi_-$, because it appears that these boundary conditions are required to describe the emergent property of time asymmetry with solid mathematics.

\section{Theory of Particle Resonances}

Assuming the resonance is isolated, the amplitude for resonance scattering has a complex pole which can be parameterized by the mass $M$ and width $\Gamma$.  However, there are several ways of achieving this parameterization (see \cite{npb00} for a review of this applied to the Z-boson).

In the on-mass shell renormalization approach, the pole in the resonance amplitude is derived as the pole in the renormalized propagator for the resonance state and takes the form
\begin{equation}\label{amp1}
a_R(\sm)\propto \frac{1}{\sm - M^2 + i\sqrt{\sm}\Gamma(\sm)},
\end{equation}
where $\Gamma(\sm)$ is a function for the width that depends on the center-of-mass energy squared $\sm$.  In the case of the Z-boson, where the decay products have only a small fraction of the the Z-boson mass, the form is chosen from phase space considerations
\begin{equation}
\Gamma_Z(\sm) = \frac{\sqrt{\sm}}{M_Z}\Gamma_Z.
\end{equation}
The parameterization of the resonance amplitude by the on-mass shell renormalization approach has been shown to be arbitrary and gauge dependent~\cite{arb}.

An alternate approach begins with the association of the resonance to a pole in the analytically continued S-matrix.  The partial amplitude with angular momentum corresponding the resonance can be broken into a non-resonant part and a pole term with the form
\begin{equation}\label{amp2}
a_R(\sm)\propto \frac{1}{\sm - \sm_R},
\end{equation}
where $\sm_R$ is a complex number indicating the location of the pole.  The complex number $\sm_R$ can be parameterized by the real mass and width in many ways, the most common being
\begin{equation}
\sm_R = M_\rho^2 - i\Gamma_\rho M_\rho.
\end{equation}
An alternate parameterization which is analogous in form to the non-relativistic pole parameterization is
\begin{equation}\label{para}
\sm_R = (M_R - i\Gamma_R)^2.
\end{equation}

How should one choose between these three forms?  All result in the same functional form for the cross section, just with different values for mass and width.  Already mentioned are the theoretical problems with (\ref{amp1}), and there are phenomenological reasons, too; form (\ref{amp2}) seems to give more consistent results for wide resonances across different decay channels~\cite{consist}.  Another distinction between the forms (\ref{amp2}) and (\ref{para}) is that only for (\ref{para}) is the definition of width consistent with the Weisskopf-Wigner relation~\cite{npb00}.  This connection is established via the relativistic Gamow ket, discussed in more depth below.  The relativistic Gamow get has a complex mass with width $\Gamma$ and exponentially decays under translations into the forward light with a rest frame lifetime that is exactly $\tau=\hbar/\Gamma$.

That the Weisskopf-Wigner should be valid is a theoretical bias that emerges from the perspective of the introduction:  unstable states, both decaying states and resonances, are representations of the Poincar\'e semigroup, i.e., elementary systems except for their instability.  This amounts to the opinion that unstable particles should be as similar in representation as possible to stable particles, except for non-zero width as an emergent property.  With that in mind, we turn to how stable particles are represented so we can see how to extend the notion to the relativistic Gamow ket.

\section{UIRs of the Poincar\'e Group}

The Poincar\'e group is the semidirect product of the orthogonal transformations in Minkowski space, $\Lambda\in\mathrm{O}(1,3)$ with the translations in $a\in\mathbb{R}^4$.  The restricted (or proper, ortho\-chron\-ous) Poincar\'e group results from the restriction of $\Lambda$ to unit determinate, i.e. the Lorentz group $\Lambda\in\mathrm{SO}(1,3)$, hence excluding time and space reflections.

For quantum mechanics, we are concerned with representations of a symmetry group on the spaces of states $\{\phi\}=\Phi_+$ and observables $\{\psi\}=\Phi_-$.  Probabilities to find a state in an observable should then respect the symmetry:
\begin{equation}
|\left\langle \phi|\psi\right\rangle|^2 = |\left\langle U(\Lambda, a)\phi|U(\Lambda, a)\psi\right\rangle|^2,
\end{equation}
where $U(\Lambda,a)$ is the unitary representation of the group element.  Since all that is measurable is probability, and since probability is proportional to the norm square of the amplitude, the group composition law can only be verified up to a phase:
\begin{equation}
U(\Lambda, a)U(\Lambda', a') = \omega U(\Lambda\Lambda', a + \Lambda a'),
\end{equation}
where $|\omega| = 1$.  Wigner showed~\cite{wigner39} that for the Poincar\'e group this phase can be reduced to $\omega=\pm 1$, and removed entirely if, instead of considering the Lorentz transformation, one considers its covering group $\mathrm{SL}(2,\mathbb{C})$, the group of complex two-by-two matrices with unit determinate.  So, to analyze the consequences of Poincar\'e symmetry for quantum mechanical systems, one studies the projective representations (representations up to a phase) and one is lead to the quantum mechanical Poincar\'e group, the semidirect product of $\mathrm{SL}(2,\mathbb{C})$ with the translations $\mathrm{T}_4$.  The result of considering the covering group is that it includes the half-odd-integer spin representations.  For simplicity, we will still notate elements of the restricted Poincar\'e group $(\Lambda, a)$ although this corresponds to two different elements of the covering group, i.e., the quantum mechanical Poincar\'e group.

Following the construction of Wigner~\cite{wigner39}, a special case of the general method of induced representations~\cite{mackey}, UIR spaces are labeled ${\rm UIR}(\sm, j)$ by a fixed pair of eigenvalues of the Casimir invariants of the Poincar\'e Lie algebra.  Poincar\'e transformations leave UIRs invariant, i.e. if $\phi\in{\rm UIR}(\sm, j)$, then $U(\Lambda,a)\phi\in{\rm UIR}(\sm, j)$.  Because of the properties of the Lie algebra of the Poincar\'e group, Wigner and many others have found it natural to identify $(\sm= m^2, j)$ with the invariant mass squared and spin of a single, interaction-free particle and therefore to associate the ${\rm UIR}(\sm, j)$ with the state (and observable) space of that particle.

For later use, we give a description of the basis used to decompose the UIR associated with mass squared $\sm$ and spin $j$ and positive energy.  A standard choice for the complete set of commuting observables (CSCO) is made, $\{M^2, J^2, P_i, S_3(P)\}$, where $P_i$ are the spatial components of the four momentum operator and $S_3(P)$ is the spin in the particle's rest frame on the 3-axis (which can be constructed from the generators of the Poincar\'e transformations, but the exact form is not required here).

Other choices for the CSCO can be made~\cite{polyzou}, but the choice
\[
\{M^2, J^2, {\bf P}, S_3(P)\}
\]
leads to the basis called the Wigner basis for the expansion of the ${\rm UIR}(\sm, j)$.  Later we will make a slight modification and use $\hat{P}_i = P_i/M$, the spatial components of the 4-velocity operators, but that will only change the normalization for the basis kets.  If the Hilbert space is chosen for the the UIR, then elements $\phi\in\mathcal{H}(\sm,j)$ can be realized as a direct product of Lebesgue square integrable functions of the momentum and Lebesgue square summable over the spin components.

If additionally the realizations of the elements of the UIR are chosen to be elements of the Schwartz space of ``well-behaved'' functions of the momentum $\phi\in\Phi(\sm,j)$, then improper eigenvectors, or Dirac eigenkets, of this CSCO $\kt{\wig}$ are elements of the linear topological dual $\Phi^\times$ of $\Phi$~\cite{bohmrhs} and have the following properties:
\begin{eqnarray}
M^2\kt{\wig} &=& \sm\kt{\wig}\nonumber \\
J^2\kt{\wig} &=& j(j+1)\kt{\wig}\nonumber \\
{\bf P}\kt{\wig} &=& {\bf p}\kt{\wig}\nonumber \\
S_3({\bf P})\kt{\wig} &=& \xi\kt{\wig},
\end{eqnarray}

We choose a relativistically invariant normalization,
\begin{equation}\label{wignorm}
\bk{{\bf p}',\xi'[\ms,j]}{\wig} = 2 \ms p_0\delta^3({\bf p'} - {\bf p})\delta_{\xi'\xi},
\end{equation}
giving the form to the expansion of a vector $\phi\in\Phi$ with invariant measure
\begin{equation}
\phi = \frac{1}{\ms}\sum_\xi \int \frac{d^3{\bf p}}{2 p_0} \kt{\wig}\bk{\wig}{\phi}.
\end{equation}

For $\phi\in\Phi(\sm,j)$ and $\kt{wig}\in\Phi^\times$, the Poincar\'e transformations then have the form
\begin{subequations}\label{wigtrans}
\begin{eqnarray}
U(\Lambda, a)\phi_\xi( p) &=& \bk{\wig}{U(\Lambda, a)\phi}\nonumber \\
&=& \br{\wig}U(\Lambda, a)\kt{\phi}\nonumber \\
&=& e^{ip\cdot a}\sum_{\xi'}D_{\xi'\xi}^j(W(\Lambda, \Lambda^{-1}p))\phi_{\xi'}({\bf \Lambda^{-1}p})\end{eqnarray}
or equivalently 
\begin{equation}
U(\Lambda, a)\kt{\wig} = e^{-i(\Lambda p)\cdot a}\sum_{\xi'}D_{\xi'\xi}^{j}(W(\Lambda, p)) \kt{{\bf \Lambda p},\chi'[\ms,j]},
\end{equation}
\end{subequations}
where $D_{\xi'\xi}^j(R)$ is the $2j+1$ dimension representation of the quantum mechanical spatial rotation $R\in{\rm SU}(2)$ and $W(\Lambda, p)\in{\rm SU}(2)$ is called the Wigner rotation. Its exact form in terms of representative boosts is related to the choice of $S_3({\bf P})$ for the spin degeneracy operator~\cite{kummer}. Using $S_3({\bf P})$, the rotation $W(\Lambda, p)$ has the form
\begin{equation}
W(\Lambda, p) = L(\Lambda p)^{-1}\Lambda L(p) 
\end{equation}
where $L(p)$ is a representative element of the left coset space ${\rm SO}(1,3)/{\rm SO}(3)$ (or considering the quantum mechanical Poincar\'e group ${\rm SL}(2,\mathbb{C})/{\rm SU}(2)$).  For our choice of $S_3({\bf P})$, the $L(p)$ are the standard boosts that take a vector from the rest frame to momentum $p$ without rotation.  Other choices of representants are connected with alternate choices for the spin degeneracy quantum number and lead to alternate bases, such as the helicity basis.

We will also consider the case where the standard representation for the parity $P$ and charge parity operator $C$ are represented by unitary linear operators $U_P$ and $U_C$.  These symmetries, while not respected by the weak interaction, are held by the strong and electromagnetic interactions and therefore may be useful for the analysis of some resonances.  Particles are assumed to have well defined parity $\pi=\pm 1$
\begin{equation}
U_P\kt{{\bf p}, \xi [\sm, j, \pi]} = \pi \kt{-{\bf p}, \xi [\sm, j, \pi]}.
\end{equation}
If the particle represented by the UIR is also an eigenstate of charge parity, then $U_C$ has the action
\begin{equation}
U_C\kt{{\bf p}, \xi [\sm, j, \pi, \eta]} = \eta \kt{{\bf p}, \xi [\sm, j, \pi, \eta]},
\end{equation}
where $\eta=\pm 1$.  If they are not eigenstates, then $U_C$ has the action
\begin{equation}
U_C\kt{{\bf p}, \xi [\sm, j, \pi, n]} = \eta \kt{{\bf p}, \xi [\sm, j, \pi, \bar{n}]},
\end{equation}
where $n$ are the eigenvalues of operators like charge that do not commute with $U_C$ and $\bar{n}$ are their conjugate values; $\eta$ only has physical significance for the case of CP eigenstates $n=\bar{n}$.  We will work in the projective representation of the extended Poincar\'e group where $(U_C U_P)^2 = \pm 1$, where the plus is for bosons and the minus for fermions~\cite{goldberg}.

\section{Relativistic Gamow Ket and the \\ Clebsch-Gordon Technique}

Here we give a brief outline of the definition and derivation of the relativistic Gamow vector and how is partially answers some of the questions raised in the introduction.  The goal is to generalize the UIR's of the Poincar\'e group ${\rm UIR}(\sm, j)$ to complex mass squared $\sm \rightarrow \sm_R = (M - i\Gamma/2)^2$.

Consider the resonant scattering amplitude from the in-state $\phi^{\rm in}$ to the out-observable $\psi^{\rm out}$:
\begin{equation}\label{scamp}
(\psi^{\rm out}, S \phi^{\rm in}) = (\Omega^-\psi^{\rm out}, \Omega^+\phi^{\rm in})=(\psi^-,\phi^+),
\end{equation}
where $S$ is the S-matrix operator $S=(\Omega^-)^\dag\Omega^+$ and $\Omega^\pm$ are the M{\o}ller wave operators.  The in-state $\phi^{\rm in}$ and out-observable $\psi^{\rm out}$ are collections of particles, for example $\phi^{\rm in}$ might be the two particles in the beams that collide to form the resonance or unstable system and $\psi^{\rm out}$ are the decay product particles.  Treating the in- and out-particles as stable, then the in-state and out-observables are vectors in the direct product spaces of the UIRs for each particle in the collection which satisfy the boundary conditions $\phi^+\in\Phi_-$ and $\psi^-\in\Phi_+$.  For the analysis to proceed, we must decompose these direct product states into elementary systems, i.e., into a direct sum of UIRs.  This technique is called relativistic partial wave analysis or the Clebsch-Gordon technique for the Poincar\'e group.

In general, the direct product of two UIRs of the Poincar\'e group can be decomposed into a direct sum of UIRs:
\begin{equation}
{\rm UIR}(\sm_1, j_1)\otimes{\rm UIR}(\sm_2, j_2) = \sum_{j,\nu}\int\, d\mu(\sm)UIR(\sm,j)^\nu,
\end{equation}
where $d\mu(\sm)$ is some integral measure over the mass squared variable labeling the direct sum UIRs, $j$ is the spin of the direct sum UIRs, and $\nu$ is a set of degeneracy parameters arising from the coupling scheme used.  This is in analogy to the Clebsch-Gordon technique for the rotation group, where two spin states are coupled into a total angular momentum state:
\begin{equation}
\kt{\xi_1,j_1}\otimes\kt{\xi_2,j_2} = \sum_{j=|j_1 - j_2|}^{j_1 + j_2}\sum_{\xi=-j}^{+j}\bk{\xi, j}{\xi_1\xi_2,j_1 j_2}\kt{\xi,j}
\end{equation}
and $\bk{\xi, j}{\xi_1\xi_2,j_1 j_2}$ is the Clebsch-Gordon coefficient (CGC) for the rotation group (there is no $\chi$ for this case).

Each of the UIR in the sum can be decomposed in its own Wigner 3-momentum spin basis which we will denote as $\kt{{\bf p}, \xi [\sm,j,\alpha_12] (\nu)}$.  These transform irreducibly in the sense of (\ref{wigtrans}).  For future equations, we establish their normalization so that the basis kets from different UIRs of the direct sum are orthogonal, as well as the standard momentum and spin component orthogonality:
\begin{equation}
\bk{{\bf p} \xi [\ms j l s\alpha_{12}] (\nu)}{{\bf p}' \xi' [\ms' j' \alpha_{12}] (\nu')} = 2 p_0 \ms^2 \delta^3(\vp - \vp')\delta_{\xi \xi'}\delta(\ms - \ms')\delta_{jj'}\delta_{\nu\nu'}.
\end{equation}
With this choice, we have
\begin{eqnarray}
\kt{{\bf p}, \xi [\sm,j,\alpha] (\chi)} &=& \frac{1}{\ms_1\ms_2}\sum_{\xi_1\xi_2}\int \frac{d^3{\bf p}_1 d^3{\bf p}_2}{4(p_1)_0 (p_2)_0}  \kt{{\bf p}_1 \xi_1 {\bf p}_2 \xi_2 [\alpha]}\nonumber\\
&& \times \bk{{\bf p}_1 \xi_1 {\bf p}_2 \xi_2 [\alpha]}{{\bf p}, \xi [\sm,j,\alpha] (\nu)},
\end{eqnarray}
and
\begin{eqnarray}
\kt{{\bf p}_1 \xi_1 {\bf p}_2 \xi_2 [\alpha]} &=& \sum_{j\nu}\int_{(m_1 + m_2)^2}^\infty\frac{d\sm}{\ms^2} \sum_\xi \frac{d^3{\bf p}}{2p_0} \kt{{\bf p}, \xi [\sm,j,\alpha] (\nu)}\nonumber\\
&&\times \bk{{\bf p}, \xi [\sm,j,\alpha] (\nu)}{{\bf p}_1 \xi_1 {\bf p}_2 \xi_2 [\alpha]}.
\end{eqnarray}
This choice of normalization and integration measure over the mass is somewhat non-standard, but is useful because then both the direct product and direct sum kets have the same dimensional units and the CGCs are unitless.

As an example, using the normalization and basis Dirac kets we have chosen and the spin-orbit angular momentum coupling scheme~\cite{joos,macfarlane}, the CGC between a single particle UIR in its rest frame $p_r= (\sqrt{\sm}, {\bf 0})$ and the direct product of two single particle UIRs with zero spin and equal mass $\sqrt{\ms_0}$ is
\begin{eqnarray}
\bk{{\bf p}_1 {\bf p}_2 [\alpha]}{{\bf p}_r \xi [\sm j \alpha] (l s)} &=& 4 \left[\frac{4\sm}{\sm - 4\ms_0}\right]^{1/4}\ms_0\ms^{3/2} \delta^3({\bf p}_1 + {\bf p}_2) \delta(\sm - 4(p_1)_0^2) \nonumber \\
 && \times \delta_{s,0}\delta_{l,j}Y_{j \xi}(\hat{{\bf p}}_1).
\end{eqnarray}
In this equation the degeneracy parameters $\nu$, now identified as $s$ and $l$, are established by the spin-orbit angular coupling scheme used.  The half-integer $s$ is the results from coupling the spins of the two particles (zero for this case) and $l$ is the integer orbital angular momentum in the rest frame. Then $s=0$ and $l$ couple into the total angular momentum $j=l$.  In general, various combinations of $s$ and $l$ give a particular $j$ and therefore a particular UIR with mass $m$ and angular momentum $j$ appears several times in the reduction of the direct product. The spherical harmonic $Y_{j \xi}(\hat{{\bf p}}_1)$ is a function of the unit direction vector $\hat{{\bf p}}_1$, which points in the direction of the relative 3-momentum in the center of mass frame.

If we also want to consider the action of $U_P$ and $U_C$, we need also to have 
the Clebsch-Gordon coefficients for the full Poincar\'e group.  These results have recently been derived~\cite{licata} and reproduce the standard relations achieved through the standard techniques~\cite{weinberg} in an easy, manifestly relativistic way.  For example, returning to the same case as above, if the equal mass particles are a particle-antiparticle pair, then the CGC becomes
\begin{eqnarray}
\bk{{\bf p}_1 {\bf p}_2 [\alpha \pi_1 \pi_2]}{{\bf p}_r \xi [\sm j \alpha \pi \eta] (l s)} &=& 4 \left[\frac{4\sm}{\sm - 4\ms_0}\right]^{1/4}\ms_0\ms^{3/2}\nonumber\\
 \delta^3({\bf p}_1 + {\bf p}_2) \delta(\sm - 4(p_1)_0^2)\delta_{s,0}\delta_{l,j}\nonumber\\
  && \times Y_{j \xi}(\hat{{\bf p}}_1)\delta_{\pi_1\pi_2(-)^l,\pi}\delta_{(-)^{l+s},\eta}  
\end{eqnarray}
where $\pi$ is the parity and $\eta$ is the charge parity of the direct sum ${\rm UIR}(\sm, j)^\nu$.

With these CGCs, we can return to the scattering amplitude (\ref{scamp}) where now the in-states and out-observables can be expressed in terms of the direct sum basis, i.e. they can be decomposed into a superposition of elementary systems.  The interacting kets in the direct sum basis can be constructed formally by the M{\o}ller wave operators,
\begin{equation}
\kt{{\bf p}, \xi [\sm,j,\alpha] (\nu){}^\pm} = \Omega^\pm\kt{{\bf p}, \xi [\sm,j,\alpha] (\nu)},
\end{equation}
and a complete set of these elementary systems can be inserted to expand the in-state and out-observable.  Note that
\begin{equation}
\bk{{\bf p}, \xi [\sm,j,\alpha] (\nu){}^+}{\phi^+} = \bk{{\bf p}, \xi [\sm,j,\alpha] (\nu)}{\phi^{in}}\in\{{\rm suitable\ realization\ of}\ \Phi_- \}
\end{equation}
and
\begin{equation}
\bk{{\bf p}, \xi [\sm,j,\alpha] (\nu){}^-}{\psi^-} = \bk{{\bf p}, \xi [\sm,j,\alpha] (\nu)}{\psi^{out}}\in\{{\rm suitable\ realization\ of}\ \Phi_+ \},
\end{equation}
where the suitable realization is provided in \cite{rgv}.

Briefly summarizing known results (see \cite{rgv} for more details), from invariance principles, the S-matrix will be diagonal in $\sm$ and $j$ and so the partial amplitude containing the corresponding to the resonance angular momentum $j_R$ can be considered.  The pole term can be separated from the non-resonanant background term and by contour integration we can define the relativistic Gamow ket from the pole term.  An expression for the relativistic Gamow ket in the rest frame then becomes
\begin{equation}\label{rgkdef}
\kt{\hat{{\bf p}}_r, \xi [\sm_R,j_R,\alpha] (\nu_R){}^-} = \frac{1}{2\pi}\int_{-\infty_{II}}^{+\infty}\frac{d\sm}{\sm-\sm_R}\kt{\hat{{\bf p}}_r, \xi [\sm,j_R,\alpha] (\nu_R){}^-}.
\end{equation}
Several things must be noted.  Here, for reasons discussed below, we have surreptitiously switched to the 4-velocity basis, where $\hat{p}=p/\sqrt{\sm}=(\hat{p}_0, {\bf \hat{p}})= (\gamma, \gamma{\bf v})$ and $i=\{1,2,3\}$.  The integral is taken on the second sheet of the S-matrix analytically continued in $\sm$.  The relativistic Gamow ket can be thought of as the analytic extension of the out-going Lippmann-Schwinger ket $\kt{\hat{{\bf p}}_r, \xi [\sm,j_R,\alpha] (\nu_R){}^-}\in\Phx_+$, and the time asymmetry comes from the analyticity requirements.

To summarize the important points from this presentation, what are the mathematical costs of generalizing the UIR of Poincar\'e group with real $\sm$ to something with complex $\sm_R = (M - i\Gamma/2)^2$?  And are these costs or profits?
\begin{itemize}
\item By taking $\sm \rightarrow \sm_R = (M - i\Gamma/2)^2$, we no longer have a UIR of the Poincar\'e, but a new object.  It is an irreducible representation IR (no longer unitary) of the Poincar\'e semigroup ${\rm IR}(\sm_R,j_R)^{\nu_R}$ labeled by the complex mass, resonance spin and internal parameters. This is the Poincar\'e group such that the set of $(\Lambda, a)$ are restricted to only those translations where $a_\mu a^\mu\geq 0$~\cite{pra02,rgv}.  
\item By making the mass complex, the momentum becomes complex. However, by considering a ``minimally complex'' representation of the ${\rm IR}(\sm_R,j_R)^{\nu_R}$, the 4-velocity will remain real.  Except for the restriction to the semigroup and the necessary normalization changes required for working with the 4-velocity eigenkets, the defining transformation representation (\ref{wigtrans}) has the same form.  And from (\ref{wigtrans}) for the Poincar\'e group, the Weisskopf-Wigner relation can be proved to hold exactly for the relativistic Gamow ket~\cite{rgv}.  This is equivalent to working with the point form of dynamics~\cite{polyzou}.
\item As mentioned before, for all this to work out we must work with elements of some spaces larger than the Hilbert space that contain the Lippmann-Schwinger kets and the relativistic Gamow ket.  The out-observable must be restricted to $\Phi_+$ and the in-state to $\Phi_-$, but this choice is equivalent to choosing causal boundary conditions for the scattering experiment~\cite{pra02}, and so this is not a cost at all.
\end{itemize}

\section{Summary and Speculation}

As a conclusion, we consider a specific case and see what questions have been answered and what questions have been raised.  The $\Upsilon(4S)$ resonance, an excited bound state of a $b\bar{b}$ quark pair, shows up in the cross section of $e^+e^- \rightarrow {\rm hadrons}$ scattering.  Its primary decay products (nearly 100\%) are the particle-antiparticle pairs $B^0\bar{B}^0$ and $B^+B^-$.  Setting our clock so it is created at $t=0$, and calling $\tau$ the lifetime of the $\Upsilon(4S)$, we might picture the following sequence of events if we wanted to adhere to the Wigner perspective of the introductory quotation.
If we could observe the $\Upsilon(4S)$ at times $t\ll \tau$, we might imagine it would appear stable, and thus be approximately represented as an element of ${\rm UIR}(M_\Upsilon^2, j_\Upsilon=1)$.  After a long time, the $\Upsilon(4S)$ will have decayed into $B^0\bar{B}^0$ or $B^+B^-$, and therefore be expressible in terms of the irreducible decomposition of a superposition (or a mixture) of ${\rm UIR}(M_{B^0}, j_{B^0}=0)\otimes{\rm UIR}(M_{\bar{B}^0}, j_{\bar{B}^0}=0)$ and ${\rm UIR}(M_{B^+}, j_{B^+}=0)\otimes{\rm UIR}(M_{B^-}, j_{B^-}=0)$.  Although the B-mesons are unstable themselves, they are very long-lived compared to the $\Upsilon(4S)$ an again could be treated as approximately stable.  Thus in these two extremes the Netwon-Wigner idea that any state can be decomposed into elementary systems seems a reasonable approximation, although this is not rigorously mathematical.

But what about when $t\approx\tau$, when the decaying is taking place?  How does the transition from a single UIR to a mixture or superposition of two other direct products of UIRs take place?  The relativistic Gamow ket gives us some hint.  The ket for the $\Upsilon(4S)$ at rest should be either a superposition or a mixture of the kets defined from (\ref{rgkdef}),
\begin{equation}
\kt{\Upsilon(4S)\ {\rm at}\ {\rm rest}} = \kt{\hat{{\bf p}}_r, \xi [(M_\Upsilon - i\Gamma_\Upsilon/2)^2,j_R=1,B^0\bar{B}^0] (s=0, l=1){}^-}
\end{equation}
and
\begin{equation}
\kt{\Upsilon(4S)\ {\rm at}\ {\rm rest}} = \kt{\hat{{\bf p}}_r, \xi [(M_\Upsilon - i\Gamma_\Upsilon/2)^2,j_R=1,B^+B^-] (s=0, l=1){}^-}.
\end{equation}
In other words, the ket to represent the $\Upsilon(4S)$ is an element of
\[
{\rm IR}(M_\Upsilon - i\Gamma_\Upsilon/2)^2,j_R=1)^{\nu_R},
\] which is a subset of the dual Hardy space $\Phx_+$ and this relativistic Gamow ket can be constructed from the UIRs of the decay products and has the quantum numbers of the decaying state.

These three representation spaces are approximations of the real spaces and a full description which requires solving the Lippmann-Schwinger equations and M{\o}ller wave operators to construct and connect the real spaces.  The question of how to construct interacting states from elementary systems is only formally complete.  Nonetheless, I hope this sheds a little light on the underlying emergent property of the Poincar\'e semigroup and partially answers the question of how to represent unstable states by decomposing them into elementary systems.

As a final comment, the perspective of reducing particle phenomena to elementary systems seems to have gone somewhat out of fashion since the rise of gauge field theory.  I believe there are many interesting questions currently unanswerable with the techniques of perturbative gauge field theory that may be more tractable from this perspective.  Two examples are the infraparticle problem~\cite{schroer} and the Blum-Saller non-minimally complex representations of the Poincar\'e group with complex mass~\cite{blumsaller}.  To follow such a program, the properties of  the Clebsch-Gordon coefficients for the Poincar\'e group are required and there are many open questions there.

\section{Acknowledgements}

I would like to thank the organizers.  The support of the NSF though the Young Researchers Travel Grant is also gratefully acknowledged.


\begin{thebibliography}{99}

\bibitem{newtonwigner} T.D.~Newton and E.P~Wigner, Rev.\ Mod.\ Phys. {\bf 21}, 400 (1949).

\bibitem{PDG} K. Hagiwara et al., Phys.\ Rev.\ D {\bf 66}, 010001 (2002).


\bibitem{blumsaller} W.~Blum, H.~Saller, Eur.\ Phys. J.\ C. {\bf 28}, 279 (2003).


\bibitem{ajp03} N.L.~Harshman, Am.\ J.\ Phys. {\bf 71}, 984 (2003).


\bibitem{PDGMC}
\begin{verbatim}http://pdg.lbl.gov/rpp/mcdata/mass_width_02.mc\end{verbatim}

\bibitem{atherton} H.W.~Atherton et al., Phys. Lett. {\bf 15B}, 81 (1985).

\bibitem{gamow} G.~Gamow, Z.\ Phys.\ {\bf 51}, 204 (1928).

\bibitem{khalfin} L.A. Khalfin, JETP Lett. {\bf 15}, 388 (1972).

\bibitem{hegerfeldt} A general result by Hegerfeldt
shows that the transition probability between two Hilbert space vectors
is either \emph{different} from zero for all
times $-\infty <t <\infty$ if it is different from zero at times
$t\geq 0$ (which contradicts causality) or is zero at all times 
$-\infty <t <\infty$ (which means there is no decay that starts at a finite
time $t=0$).  
In other words, for resonances the decay probability has to be
zero for (almost) 
all $t$ if it is 
zero in any time interval $\Delta t$ before the time $t_0$ at which the
decaying state has been prepared (e.g. before the atom
has been excited into a 
resonance state).  See G.C.~Hegerfeldt, Phys.\ Rev.\ Lett.\ {\bf 72}, 596 (1994); \emph{Irreversibility and Causality in Quantum
Theory---Semigroups and Rigged Hilbert Space}, Vol.\ 504, Springer 
Lecture Notes in Physics,
A.~Bohm, H.-D.~Doebner,
P.~Kielanowski, [Eds.], (Springer, Berlin, 1998), p.238.

\bibitem{pra02} See also A.~Bohm, N.L.~Harshman, and H.~Walther, Phys.\ Rev.\ A {\bf 66}, 012107 (2002).

\bibitem{bohmrhs} A.~Bohm, in \emph{Studies in Mathematical Physics},
Proceedings of the Istanbul Summer Institute, (D.~Reidel Pub.\ Co., 1973); A.~Bohm, M.~Gadella, \emph{Dirac Kets, Gamow Vectors and Gel'fand Triplets}, Lecture Notes in Physics, Vol. 348, (Springer, 1989).

\bibitem{rgv} A.~Bohm, H.~Kaldass, S.~Wickramasekara, Fortsch.\ Phys.\ {\bf 51}, 569 (2003);  Fortsch.\ Phys.\ {\bf 51}, 604 (2003).

\bibitem{rhsrev} A.~Bohm, I.~Antoniou, P.~Kielanowski, Phys.\ Lett.\ A
{\bf 189}, (1994) 442.  For a review, see 
A.~Bohm, N.L.~Harshman, in \emph{Irreversibility and
Causality}, A.~Bohm, H.-D.~Doebner, P.~Kielanowski [Eds.]
(Springer, Berlin, 1998) p.~181.

\bibitem{npb00} Arno R.~Bohm and N.L.~Harshman, Nucl.\ Phys.\ B
{\bf 581}, 91 (2000).

\bibitem{arb} M.~Consoli and A.~Sirlin, in Physics at LEP, Vol.~1,
J.~Ellis and R.~Peccei [Eds.], CERN report CERN 86-02 (1986) p.~63; F.~A.~Berends, G.~Burgers, W.~Hollik and W.~van
Neerven, Phys.~Letters B {\bf 203} (1988) 177; S.~D.~Bardin, A.~Leike, T.~Riemann, M.~Sachwitz,
Phys.~Letters B {\bf 206} (1988) 539; S.~Willenbrock, G.~Valencia, Phys.~Lett.\
B {\bf 259} (1991) 373; R.~G.~Stuart, Phys.~Lett.\ B {\bf 262} (1991) 113;
{\bf 272} (1991) 353; A.~Sirlin, Phys.~Rev.~Lett.\ {\bf 67} (1991) 2127;
Phys.~Lett.\ B {\bf 267} (1991) 240.

\bibitem{consist} A.~Bernicha, G.~L{\'o}pez Castro, J.~Pestieau,
Phys.~Rev.~D {\bf 50}, 4454 (1994); A.~Bernicha, G.~L{\'o}pez Castro,
J.~Pestieau, Nucl.~Phys.~A {\bf 597}, 623 (1996).

\bibitem{wigner39} E.~P.~Wigner, Ann.\ Math.\ (2) {\bf 40}, 149 (1939).

\bibitem{mackey} G.W.~Mackey, Amer.\ J.\ Math.\ {\bf 73}, 576 (1951).

\bibitem{polyzou} B.D.~Keister, W.N.~Polyzou, in \emph{Advances in Nuclear Physics}, Volume 20, J. W. Negele and E.W. Vogt [Eds.], (Plenum, 1991), p.~226; W.N.~Polyzou, in \emph{Proceedings of the XIII European Conference in Few-Body Problems in Physics}, Few-Body Systems Supplement {\bf 14}, 147 (2003); W.N.~Polyzou, Phys. Rev. C {\bf 68}, 015202 (2003).

\bibitem{kummer} M.~Kummer, J.\ Math.\ Phys.\ {\bf 7}, 997 (1966).

\bibitem{goldberg} H.~Goldberg, N.\ Cim.\ {\bf 60}, 509 (1969); R.~Scurek, Am.\ J.\ Phys.\ {\bf 72}, 638 (2004).

\bibitem{joos} H.~Joos, Fortsch.\ Phys.\ {\bf 10}, 65 (1962).

\bibitem{macfarlane} A.J.~Macfarlane, Rev.\ Mod.\ Phys. {\bf 34}, 41 (1962).

\bibitem{licata} N.L.~Harshman, N.~Licata, \texttt{hep-ph/0407299}.

\bibitem{weinberg} Steven Weinberg, \emph{The Quantum Theory of Fields}, Volume 1, (Cambridge University Press, 1995).

\bibitem{schroer} B.~Schoer, Fort.\ Phys.\ {\bf 11}, 1 (1963); J.~Fr\"olich, G.~Morchio, F.~Strocchi, Phys.\ Lett.\ B {\bf 89}, 61 (1979); D.~Buchholz, Phys.\ Lett.\ B {\bf 85}, 331 (1986).

\end{thebibliography}
\end{document}